\documentclass[conference]{IEEEtran}
\IEEEoverridecommandlockouts
\usepackage{cite}
\usepackage{amsmath,amssymb,amsfonts}
\usepackage{algorithm}
\usepackage[noend]{algpseudocode}
\makeatletter
\def\algbackskip{\hskip-\ALG@thistlm}
\makeatother
\renewcommand\IEEEkeywordsname{Keywords}
\usepackage{graphicx}
\usepackage{textcomp}
\usepackage{xcolor}
\usepackage{balance}
\usepackage{graphicx}  
\usepackage{epstopdf}  

\def\BibTeX{{\rm B\kern-.05em{\sc i\kern-.025em b}\kern-.08em
    T\kern-.1667em\lower.7ex\hbox{E}\kern-.125emX}}
\begin{document}

\title{Privacy-Enhanced Living: A Local Differential Privacy Approach to Secure Smart Home Data}

\author{
\fontsize{9}{11}\selectfont
Nazar Waheed\textsuperscript{1},
Fazlullah Khan\textsuperscript{2},
Spyridon Mastorakis\textsuperscript{3},
Mian Ahmad Jan\textsuperscript{2},
Abeer Z. Alalmaie\textsuperscript{1},
Priyadarsi Nanda\textsuperscript{1}\\
\fontsize{9}{11}\selectfont
\textsuperscript{1}\textit{School of Electrical and Data Engineering, University of Technology Sydney, Sydney, Australia}\\
\fontsize{9}{11}\selectfont
\textsuperscript{2}\textit{Department of Computer Engineering, Abdul Wali Khan University, Mardan, Pakistan}\\
\fontsize{9}{11}\selectfont
\textsuperscript{3}\textit{Computer Science and Engineering Department, University of Notre Dame, USA}
}
\IEEEoverridecommandlockouts
\IEEEpubid{\makebox[\columnwidth]{979-8-3503-4647-3/23/\$31.00~\copyright2023 IEEE \hfill} \hspace{\columnsep}\makebox[\columnwidth]{ }}

\maketitle
\IEEEpubidadjcol
\begin{abstract}

The rapid expansion of Internet of Things (IoT) devices in smart homes has significantly improved the quality of life, offering enhanced convenience, automation, and energy efficiency. However, this proliferation of connected devices raises critical concerns regarding security and privacy of the user data. In this paper, we propose a differential privacy-based system to ensure comprehensive security for data generated by smart homes. We employ the randomized response technique for the data and utilize Local Differential Privacy (LDP) to achieve data privacy. The data is then transmitted to an aggregator, where an obfuscation method is applied to ensure individual anonymity. Furthermore, we implement the Hidden Markov Model (HMM) technique at the aggregator level and apply differential privacy to the private data received from smart homes. Consequently, our approach achieves a dual layer of privacy protection, addressing the security concerns associated with IoT devices in smart cities.
\end{abstract}

\begin{IEEEkeywords}
Smart Homes, Internet of Things, Local Differential Privacy, Security, Hidden Markov Chain.
\end{IEEEkeywords}

\section{Introduction}

Advancements in software and hardware have driven the expansion of information and communication technologies (ICT), playing a crucial role in smart city development. By incorporating ICT into urban operations, cities become more efficient and adaptable, leading to the prevalent term "smart city." These urban environments leverage ICT and other strategies to enhance residents' quality of life, catering to the needs of present and future generations across social, environmental, and economic dimensions.

The Internet of Things (IoT) is an essential component in the development of smart cities, acting as their backbone. Smart cities are made feasible through the use of IoT, which includes smart sensors, smartphones, radio-frequency identification (RFID), and smart meters as central elements of the IoT framework. The IoT framework comprises various modules, such as electronics, firmware, networks, sensors, and software. Wireless devices, including sensors, displays, actuators, and home appliances, are connected through IoT, enabling a large amount of data to be generated and exchanged among devices and the Internet to achieve ubiquitous interconnectivity. IoT devices have propelled a data explosion, transferring vast amounts of data to the cloud for real-time processing in applications such as electronic healthcare systems, vehicular ad hoc networks (VANETs), and smart homes \cite{zhang2011home}. In IoT networks, sensor data is collected from various applications, and different sensor device data is analyzed using deep learning approaches \cite{zhao2019differential}. Sundaravadivel et al. \cite{sundaravadivel2018smart} proposed a system based on IoT called smart-Log, which identifies nutritious food items for children using deep learning. IoT has numerous applications, such as VANET and smart homes \cite{denning2013computer}, and electronic healthcare systems used for real-time data processing.

Smart Home is a vital IoT application that uses connected devices to make our lives more efficient and convenient. Smart homes provide security to homeowners and can be controlled remotely, offering comfort and security \cite{denning2013computer}. Sensor-collected data can be used for home activity prediction within smart homes \cite{zhang2018enabling}, smart healthcare for patient treatment \cite{muhammad2018edge}, disorder assessment, and smart city pedestrian monitoring \cite{lwowski2017pedestrian}. In smart cities, data is exchanged among smart homes. To participate in smart cities, people must feel secure and protected. Security and privacy protections are essential, especially when data is transferred from one area to another with multiple parties having access to it. Various techniques in the literature are used for data privacy, such as $k$-anonymity, $l$-diversity, $t$-closeness, and differential privacy, respectively.

This paper aims to explore differential privacy techniques for ensuring the data privacy of Smart homes. In 2006, Dwork proposed the use of differential privacy to prevent adversaries from accessing data. Differential privacy is a crucial technique for data privacy. This paper focuses on securing smart home data using local differential privacy (LDP). In smart homes, people do not want their data to be accessible to outsiders. Centralized Differential Privacy (CDP) is based on the assumption that the aggregator node will be honest, which is difficult to guarantee in real life. To preserve privacy, LDP is suggested with strong privacy guarantees \cite{mahmud2019smart, raskhodnikova2008learn}, which is an extension of differential privacy. LDP can resist adversaries with background knowledge and uses distributed randomized processes to prevent data leakage. LDP has various industrial applications, such as Google's LDP structure called "RAPPOR" \cite{dwork2006differential}, which is used in Chrome to collect user behavior data. Apple announced the use of LDP for user privacy preservation at WWDC 2016 \cite{zhao2014achieving}.

In this paper, we propose a method for sending home data to the aggregator while considering that aggregator nodes might be malicious. Therefore, LDP is applied to the real-time data of homes. The data is privatized before being sent to the aggregator. In our proposed model, an obfuscation method is applied at the aggregator side to ensure that the aggregator cannot recognize the homeowner, achieving anonymity. Our model also incorporates the hidden Markov model (HMM) concept, a widely used approach for time series data modeling with applications in various areas such as bioinformatics, speech recognition, and Internet traffic modeling. Once the model is trained, it can be used to detect anomalies by scanning for unlikely series of observations. The aggregator also employs CDP on privatized home data. Our research aims to achieve double privacy.

The contributions of this paper are as follows:

\begin{itemize}
\item We design a secure smart home data collection framework.
\item We calculate the privacy risk using a probabilistic technique based on the hidden Markov model (HMM), which computes the probabilities of smart home data.
\item We apply an obfuscation method to obfuscate high-risk data to achieve anonymity.
\item We utilize the differential privacy concept at the aggregator side to achieve double privacy.
\end{itemize}

The rest of this paper is organized as follows: Section 2 provides a literature review. Section 3 presents the problem along with a background introduction on DP, LDP, HMM theory, and Randomized Response. Section 4 describes the proposed scheme. In Section 5, the evaluation of the proposed scheme is presented followed by its limitations in Section 6. Finally, the paper is concluded and future research directions are provided in Section 7.

\section{Related Work}

The development of information and communication technology has facilitated more convenient lives for residents in smart cities. However, the transmitted data may contain sensitive information necessitating privacy protection. Various privacy techniques, including differential privacy, have been proposed by researchers to ensure data privacy. This paper specifically focuses on securing smart home data using local differential privacy.

Dwork introduced differential privacy in 2006 to prevent unauthorized access to data \cite{dwork2006differential}. Subsequent research has expanded on differential privacy in various contexts, such as battery load balancing \cite{zhao2014achieving}, cost reduction in smart meters \cite{zhang2016cost}, and edge filtering for reducing calculation and communication overhead \cite{xu2018distilling}. Privacy-preserving structures for smart homes have also been proposed, including LDP-based schemes for reducing energy consumption and preserving privacy \cite{liu2018epic} and the Differential Privacy-based Real-time Load Monitoring (DPLM) approach for concealing load values \cite{hassan2019differential}.

In a study by Wang et al. \cite{wang2018secure}, the LDP concept was employed to privatize hospital data, but it did not protect individual privacy. Our proposed system addresses this limitation by utilizing an obfuscation method to achieve individual privacy. Additionally, we apply centralized differential privacy at the aggregator level to provide dual-layer privacy protection.

\section{Preliminaries}
\subsection{Differential Privacy}
Differential Privacy, also known as Centralized Differential Privacy (CDP), is a privacy model in which data is entrusted to a third party, often a database owner. This entity receives queries and provides responses with the addition of a specific level of noise to the data to ensure privacy. Formally, differential privacy is predicated on the concept of neighboring datasets, denoted as $q$ and $q'$, which differ by only a single data point.

\textbf{Definition 1 (Neighboring Datasets)}: A randomized algorithmic function $G$ satisfies the condition of $\varepsilon$-differential privacy if, for any possible outcome $h \in \text{Range}(G)$ and any two adjacent datasets $q$ and $q'$, the following inequality holds:
\begin{equation}
    P[G(q) \in h] \leq \exp(\varepsilon) \times P[G(q') \in h] \hspace{5mm} \label{eq1}
\end{equation}

In Equation \ref{eq1}, $\varepsilon$ is the privacy parameter, which controls the privacy level of the proposed mechanism and the resulting output function $G$. $\text{Range}(G)$ denotes the range of the function. A smaller value of $\varepsilon$ is desired for achieving higher privacy, and vice versa.

\subsection{Local Differential Privacy}
Local Differential Privacy (LDP) is a more recent model for privacy, which relies heavily on local variants. The aim is to reduce trust in third-party data aggregators, collectors, or other entities by adopting a zero-trust approach. In this scenario, individuals generate locally differential private results by adding noise to their data before transmitting the scrambled information for aggregation. However, the noise in LDP is typically larger than in Centralized Differential Privacy (CDP).

LDP is used when there is no desire to trust a centralized aggregator. By applying a randomized response, the data is obscured by the data holder at the local level. The data holder then sends the concealed data to a potentially untrusted data aggregator. To formally define LDP, let $D$ be the complete dataset, and consider a randomized algorithm $T$ that takes two data tuples $a$ and $b$ as input and produces output $a^*$. $\varepsilon$-local differential privacy (or $\varepsilon$-LDP) is defined on $T$ and $\varepsilon > 0$, which is the privacy parameter, as follows:

\textbf{Definition 2 ($\varepsilon$-local differential privacy)}: A randomized algorithm $T$ satisfies $\varepsilon$-local differential privacy if and only if for output $a^*$ and two input tuples $a, b \in D$, the following inequality is satisfied: 
\begin{equation}
P[T(a) = a^*] \leq e^{\varepsilon} \times P[T(b) = a^*]\label{eq2}   
\end{equation}

In simple terms, LDP implies that the data aggregator cannot confidently determine whether the input record is $a$ or $b$ by observing the output $a^*$. LDP differs from CDP, which is defined on two neighboring databases. The two databases differ by only one record.

\subsection{Randomized Response Method}
The Randomized Response (RR) method was proposed by H. Warner et al. in 1965 \cite{dwork2006differential}. It is commonly used in Local Differential Privacy (LDP) approaches, such as RAPPOR \cite{erlingsson2014rappor}. In RR, an end user is asked a question with a binary answer, either "yes" or "no." A coin is flipped with a probability of $p$ for showing heads. To protect the end user's privacy, RR allows the end user to respond with the opposite answer when heads are shown. As a result, the data aggregator cannot confidently determine the accurate answer for a particular end user.

\textbf{Definition 3}: The RR mechanism is a mapping with $A = B$ that satisfies the following equality:

\begin{equation}
    Q(a|b) = 
\begin{cases}

\frac{e^\varepsilon}{|B| - 1 + e^\varepsilon}, & \text{if}\ a = b, \\
\frac{1}{|B| - 1 + e^\varepsilon}, & \text{if}\ a \neq b.

\end{cases}
\end{equation}\label{eq3}

Here, $Q(a|b)$ is the conditional probability, $B$ is the true dataset, $A$ is the privatized dataset, $b \in B$, $a \in A$, $|B|$ is the size of set $B$, and $\varepsilon$ is the privacy parameter.

\subsection{HMM Theory}

A Hidden Markov Model (HMM) is a sequence of random variables with the Markov property. HMM is effective in identifying network anomalies and is used in various applications. By utilizing HMM, it is possible to estimate the likelihood of an observed sequence within the home data.

\subsubsection{The Learning Problem}
The learning problem involves estimating model parameters $w = {A,B,\pi }$ from a set of observations. The set of hidden states and set of observation symbols $V$ can also be estimated.

\subsubsection{The Evaluation Problem}
In the evaluation problem, the set of observation sequence $O$ and a model $P(O/w)$ are considered. The probability $P(O/w)$ represents how well the model matches the observations.

Our research focuses on a distributed approach for solving HMM inference problems, where each household's data is kept confidential. We propose a solution that computes the HMM without compromising individual data privacy.

\section{Proposed Scheme}

\begin{figure}[!ht]
		\centering
		\begin{center}
			\includegraphics[width=3.0in,height=2.00in]{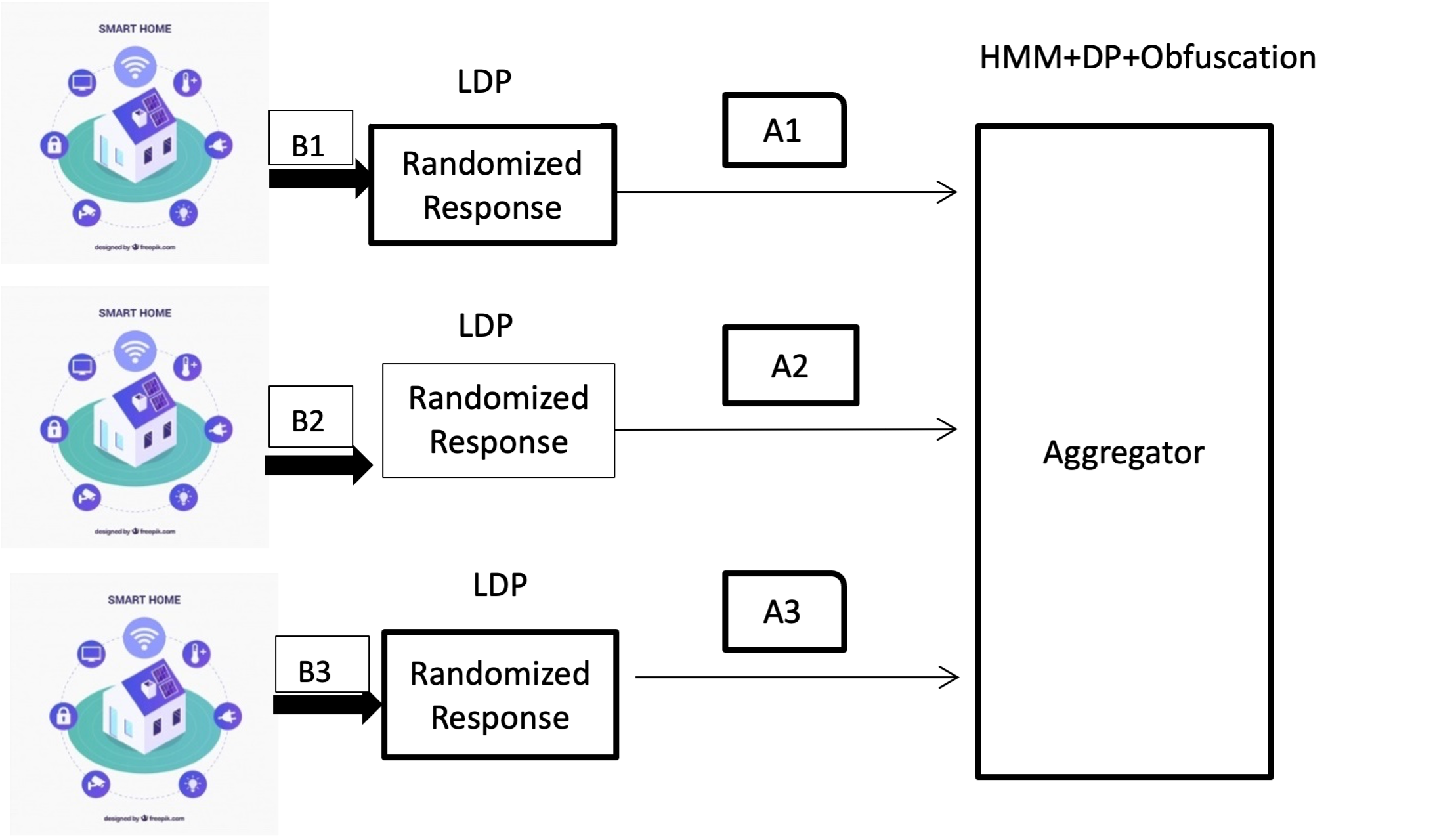}
			\vspace{-0.2cm}
			\caption{Securing Smart Homes using LDP.}
			\label{fig:SSHLDP}
		\end{center}
		\vspace{-0.4cm}
	\end{figure}

Our research paper proposes a differential privacy-based system to protect the privacy of data collected from smart homes by adding noise using the LDP approach. The RR algorithm is used at the client end to achieve privacy, followed by an obfuscation method applied at the aggregator end to ensure individual privacy. The proposed system provides double privacy protection for the client and the aggregator, respectively.
To ensure home data privacy, we use the LDP approach, where the data from smart homes is sent to the aggregator, which could be potentially malicious. The HMM model is used at the aggregator node to allow secure distributed computation without leaking data. In addition, an obfuscation method is used to achieve anonymity and prevent the aggregator from recognizing the homeowner.
Our primary objective is to achieve double privacy by applying LDP and CDP methods. The proposed system addresses the concerns of smart home individuals unwilling to have their private data leaked. Using LDP and HMM models in our system ensures that computation can be performed using private data without revealing it to unauthorized third parties as shown in Fig. \ref{fig:SSHLDP}.


\subsubsection{Approach}
The aggregator node can potentially leak the home data, so we apply the LDP concept to the data as the first line of defense. We apply the LDP algorithm, which is provided below, to ensure the privacy of the home data being sent to the aggregator node. 

\begin{algorithm} \label{Algo 1}
    \caption{LDP Algorithm}\label{algo1}
    \hspace*{\algorithmicindent} \textbf{Input:} $b$ = real data, $B$ = real dataset, $n$ = the size of $B$ \\ \hspace*{\algorithmicindent} $R$ = random number between 0 and 1 \\
    \hspace*{\algorithmicindent} $\epsilon$ = The private budget \\
    \hspace*{\algorithmicindent} \textbf{Output:} The privatized data $A$ 
    \begin{algorithmic}[1]
    
    \If {$R < \frac{e^\epsilon}{|B|-1+e^\epsilon}$} \Return b
    \Else {
        $index = R - \frac{e^\epsilon}{|B|-1+e^\epsilon} mod \frac{1}{\frac{e^\epsilon}{|B|-1+e^\epsilon}}$
        \For {$bi$ in $B$}
            \If {$i = index$ and $bi \neq b $} \Return $bi$
            \EndIf
            \If {$i = index$ and $bi = b $} \Return $bi$+1
            \EndIf    
        \EndFor
        }
    \EndIf
    
    \end{algorithmic}
\end{algorithm}

\begin{figure}[!ht]
		\centering
		\begin{center}
\includegraphics[width=3.0in,height=1.70in]{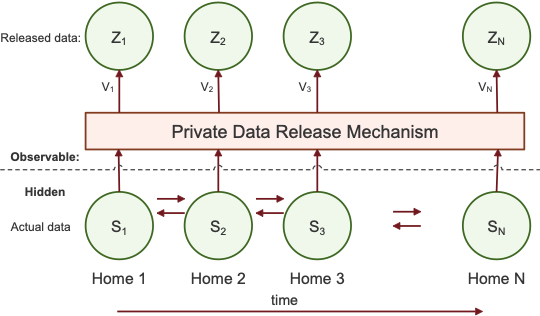}
			\vspace{-0.2cm}
			\caption{Overview of the HMM.}
			\label{fig:states}
		\end{center}
		\vspace{-0.4cm}
	\end{figure}

\paragraph{HMM at Aggregator node}
The data is forwarded to the aggregator, where the HMM is applied. Inspired from \cite{xiao2016dphmm} as shown in Fig. \ref{fig:states}, $S_1, S_2, \dots, S_N$ are the hidden states that are used to represent $Home1, Home2, \dots, HomeN$, respectively. Time is divided into $T$ slots, i.e., $t$=$\{1,2,3,...T\}$. At time t=1, the state is called $q_1$; at t=2, the state is called $q_2$; and at t=T, the state is called $q_T$, respectively. Meanwhile, $Z_1, Z_2, \dots, Z_N$ are called the visible states.

Consider an HMM with parameters ${A, B, \pi}$ as mentioned in the previous section. Observations are carried out at time intervals $0$ and $T$. Every home's data is observed in the system. The sequence of $T$ observations in which home data is observed is represented as $O_j = {O_{j1}, O_{j2}, \dots, O_{jT}}$. 



By including a null state, the HMM can be easily extended with $v$ to denote no observation. We assume that the observations of different homes are independent. Furthermore, we assume in our paper that the probability of observation for different homes' data is similar and denoted by $b_{jk}$. So,

\begin{equation}
P\left(\frac{O}{q_t}\right) = S_i = b_{jk}\label{eq4}
\end{equation}

\subsection{Using HMM for the Evaluation Problem}
We use two different HMM procedures to solve the evaluation problem: a forward procedure and a backward procedure. In the forward procedure, we need to calculate $P\left(\frac{O}{w}\right)$, which is the likelihood of  observation given the parameters $w = (\pi, A, B)$. The forward variable can be defined as: $\alpha_t(i) = P(O_1, O_2, \dots, O_t, q_t = s_i | w)$


Here, $\alpha_t(i)$ is the partial observation probability from 1 to $t$, and at this time, the state is $s_i$, given the model parameter $w$.

The complete procedure for forward HMM is shown in Algorithm 2.

\begin{algorithm}
    \caption{Forward HMM Algorithm}\label{algo2}
    \begin{algorithmic}[1]
    \State \textbf{Initialize:} $t \leftarrow 1$, $a_{ij}$, $b_{jk}$, $O_T$, $\alpha_{jt}$
    \For{Each iteration $t = 1:T $}
        \State $\alpha_{j(t)} = b_{jk} v_t \sum_{i=1}^{N} \alpha_{i(t-1)} a_{ij}$;
        \State loop until $t = T$
    \EndFor
    \State \Return $P\left(\frac{O}{w}\right) \leftarrow \alpha_{j(T)}$ for Final state
    \end{algorithmic}
\end{algorithm}

\subsection{Training HMM}
The backward procedure for backward HMM is reflected by Algorithm 3. 

\begin{algorithm}
\caption{Backward HMM Algorithm}\label{algo3}
\hspace*{\algorithmicindent} \textbf{Initialize:} $t=T$, $b_{j}T$, $a_{iJ}$, $b_{jk}$, $O^T$ \
\begin{algorithmic}[1]
\For{Each iteration $t = T-1$ to $1$ in reverse order}
\State $\beta_{i}(t) = \Sigma_{j=1}^{N} a_{ij} b_{jk} O_{t+1} \beta_j (t+1)$ \
\EndFor
\Return $P\left(\frac{O}{w}\right)$ $\leftarrow$ $\beta_{i}(1)$ for the known initial state
\end{algorithmic}
\end{algorithm}

\subsection{The Baum-Welch Algorithm}

To effectively select the parameter $w$, there is no known method to directly maximize the observed sequence probability.
We also require the following variable $\gamma_{ij}(t)$, which is the probability of being in a state $s_i$ at time $t-1$ and state $s_j$ at time $t$ given the observations:

\begin{equation}
\gamma_{ij}(t) = P(q_{t-1} = s_i, q_t = s_j | O, w) \label{eq6}
\end{equation}

The $\gamma_{ij}(t)$ can be calculated from the forward and backward variables $\alpha_i(t-1)$ and $\beta_j(t)$ as:

\begin{equation}
\gamma_{ij}(t) = \frac{\alpha_i(t-1) a_{ij} b_{jk} \beta_j(t)}{P\left(\frac{O}{w}\right)} \label{eq7}
\end{equation}

\subsubsection{Expected number of transitions from $s_i$ to $s_j$}

For a sequence $O$ at any time, it will be simply $\sum_{t=1}^{T-1} \gamma_{ij}(t)$ transitions.
Thus, it gives the expected number of transitions from state $s_i$ to state $s_j$. The total number of expected transitions from $s_i$ to any state is:
$\sum_{t=1}^{T-1} \sum_{j=1}^{N} \gamma_{ij}(t)$. Thus, when these two quantities are known, then we can update the transition probability $a_{ij}$ as:

\begin{equation}
a_{ij} = \frac{ \sum_{t=1}^{T-1} \gamma_{ij}(t)}{\sum_{t=1}^{T-1} \sum_{j=1}^{N} \gamma_{ij}(t)} \label{eq8}
\end{equation}

Similarly, we can update the emission probability $b_{jk}$ as:

\begin{equation}
b_{jk} = \frac{ \sum_{t=1}^T \xi_{jt}(t)}{\sum_{t=1}^T \sum_{k=1}^{M} \xi_{jt}(t)} \label{eq9}
\end{equation}

where, $v_t = v_k$ and $\xi_{jt}(t) = \gamma_{jt}(t)$ if $O_t = v_k$, otherwise $\xi_{jt}(t) = 0$.

The algorithm starts with arbitrary values for $a_{ij}$ and $b_{jk}$. Then, we estimate $\alpha_i(t)$ and $\beta_j(t)$ using those values. After that, we calculate $\gamma_{ij}(t)$ and update the values of $a_{ij}$ and $b_{jk}$. This process is repeated until the algorithm converges. At that point, the updated values of $a_{ij}$ and $b_{jk}$ are used to evaluate the model $w$ for any given observation sequence $O(t)$. This is how the HMM is trained.

\subsection{Obfuscation at Aggregator}

The privacy risk for home data is assessed based on the predicted privacy probability. To mitigate high privacy risks at the expense of utility loss, high-risk data is replaced with alternative data from various paths in the HMM. This process transforms the high-risk data into low-risk data. Alongside the HMM model, a list of alternative data suggestions is generated, each with their respective privacy risk and computed utility loss. The system selects a single substitute data point from this list to control the privacy risk.

\subsection{Adversarial Machine Learning}

Our obfuscation technique is vulnerable to privacy attacks since both the trained dataset and adversaries have access to the learned probabilities of the HMM. This vulnerability could allow adversaries to estimate the data by computing or guessing privacy risk values using the learned HMM probabilities and potentially compromise the privacy. The adversary may employ various methods within the HMM with high risks to deduce the data. To address this issue, we propose to incorporate the adversarial machine learning techniques into our HMM model by adding noise. This noise addition is determined by the privacy parameter $\epsilon$ and query function sensitivity $S$, and is introduced in terms of count/probabilities. Therefore, the degree of noise addition depends on both $\epsilon$ and $S$, respectively.

 \section{Evaluation}
We design a series of tasks to evaluate and validate our work. After discussing the datasets used, we present the outcomes in this section.

\subsection{Dataset}


To assess the privacy risks associated with home data and measure the effectiveness of our obfuscation method, we use two datasets in this paper: 1) Home residents' search queries related to specific illnesses, and 2) Energy appliance usage within smart homes. We collect smart home data over a two-month duration, comprising data from 40 homes. We focus primarily on cancer and diabetes as privacy-sensitive topics that may reveal user information. To extract queries related to these two topics, we first identify relevant keywords for each topic. We use the Free Keyword Tool provided by \textit{Wordstream} that utilizes the latest Google keyword API. We perform topic modeling on these keywords to obtain the most accurate and relevant terms. We then create synthetic data to simulate real-world situations. For the privacy model, we assume that two diseases are being investigated in smart homes. In dataset \textit{D}, the ratio of these diseases is: $|Cancer|:|Diabetes| = 2:4$, where $|Diabetes|$ represents the number of residents searching for information on diabetes.

\begin{table}[htbp]
\caption{Datasets used} 
\centering
{\renewcommand{\arraystretch}{1.15}
\resizebox{.9\columnwidth}{!}{%
\begin{tabular}{|c|c|c|}
\hline
 & \textbf{Search queries data related to diseases} & \textbf{Data usage of energy Appliances} \\ \hline
\textbf{Number of entries (E)} & 2000 & 45 (Total Appliances) \\ \hline
\textbf{Number of Users (U)} & 45 & 30 (Used Appliances) \\ \hline
\textbf{Number of Applications} & 10 & - \\ \hline
   &   E>100  & E >= 15 and E <= 20 \\ \hline
\textbf{Sample data} & 1200 (E), 30(U) & 22 energy appliances used in homes \\ \hline
\end{tabular}}}
\label{tab:dataset1}
\vspace{-0.3cm}
\end{table}

Table \ref{tab:dataset1} presents various diseases among residents of different homes, who seek to collect information on these diseases. However, residents are unwilling to share or reveal any health information to outsiders, as this information is sensitive and poses a high risk. Unauthorized individuals accessing or modifying the queries in the data of some homes could potentially discover a specific resident's disease status. 


The research study involved the participation of home residents who provided data for investigating specific diseases through internet searches. Upon conducting five or more queries, the residents observed a significant increase in the privacy risk percentage. We discovered that the privacy risk progressively increases with the addition of varying numbers of queries, ultimately reaching 100\% after five or more queries.


We discuss the outcomes of our study, evaluating the effectiveness and performance of a deferentially private home data obfuscation method.

\subsubsection{Obfuscation}

We analyze the results of our adversarial-resistant obfuscation method applied to sensitive home data, comparing actual and obfuscated data based on utility loss and privacy risk. While some obfuscated data achieves zero privacy risk with low utility loss, other instances result in significant alteration of the original meaning. Our findings reveal that adding noise to the HMM structure negligibly affects home data risk computation, but applying this to appliance datasets incurs considerable utility loss. We observed substantial utility loss in the Disease dataset as well. These results suggest that obfuscated data risks losing the original data's meaning.

Adding noise improves privacy and protection against adversarial attacks by securing the obfuscation method. Even with knowledge of the data, HMM probabilities and framework, adversaries struggle to discern initial privacy risk and differences between actual and obfuscated data due to noise.

\subsubsection{Efficiency}

We assessed the time efficiency by monitoring computation time as the number of data entries increased. We found that, after a certain number of queries, each query's evaluation and obfuscation time became constant due to repeated queries or constant HMM model training. New queries took longer to process. The average time for 49 cancer-related queries was 250 seconds.

\subsubsection{Throughput}

Our research demonstrates that home data privacy can be achieved using techniques like differential privacy, Local Differential Privacy (LDP), Hidden Markov Model (HMM) structures, and obfuscation methods, respectively. Using LDP ensures data privacy, as the aggregator can only access perturbed data. We applied obfuscation methods on the aggregator side and used the HMM model to compute home data' privacy risk. For high-risk data, we added noise to the HMM model to prevent adversaries from accessing the original home data. We also examined the trade-off between utility and privacy, finding that home data privacy can be achieved with a small utility cost. Our system for securing smart home data using local differential privacy is efficient and reliable.

\subsection{Discussion}
In our research, we investigated an obfuscation method for data privacy, focusing on three main aspects: evaluating home data privacy risk, semantic matching for obfuscating data entries, and resilience against adversarial machine learning. Our findings suggest that privacy risk techniques can effectively identify high-risk items, and our proposed obfuscation method efficiently safeguards against adversarial attacks. We drew several key conclusions from our study. First, privacy risk increases as more data is shared, even when user identities remain unknown, since users can be identified through their unique touch gestures and usage behaviors. Second, sharing similar data leads to heightened privacy risk. Lastly, adversaries who have knowledge of the dataset and obfuscation method can distinguish between original and obfuscated data. Incorporating differential privacy can offer protection against such attacks, but this comes at the cost of reduced utility.

\balance
\section{Limitations}

In our study, we employ a basic Hidden Markov Model (HMM) for the calculation of privacy probabilities and their corresponding privacy risk. We have not explored alternative probabilistic models, such as the Maximum Entropy Markov Model (MEMM), Gaussian distribution, or Dirichlet distribution, for comparative purposes. Our system can be extended by substituting the HMM with an equivalent alternative probabilistic model.

\section{Conclusion and Future Work}

In our research, we employ the Local Differential Privacy (LDP) technique and propose a framework for securing data collection in smart homes based on the $k$-Anonymity Randomized Response (k-RR) algorithm. Although the literature contains numerous studies on obfuscation methods addressing the privacy risks associated with internet data entries, these approaches tend to be specific to particular data types and do not universally apply. Additionally, they often neglect obfuscating high-risk data using semantically similar information.
Moreover, adversarial machine-learning techniques have yet to be thoroughly explored in the context of home data obfuscation. To address the limitations of existing methodologies, we propose a privacy-aware data obfuscation approach in our study. Through experiments utilizing home data, our results demonstrate the effectiveness of the proposed method in evaluating the privacy risk of obfuscated and original high-risk home data.
In future research, we aim to expand the applicability of our proposed scheme to additional home datasets, such as smart meter data, for monitoring residential electricity consumption. Furthermore, we envision adapting our obfuscation method into a user-centric application, potentially a browser plug-in, to enhance individual privacy protections.

\bibliographystyle{IEEEtran}
\bibliography{biblo}

\end{document}